\documentclass[12pt]{article}

\usepackage{latexsym}
\newif\ifams\amsfalse
\amstrue \ifams
 \usepackage{amsfonts}
\fi
\newcommand{\bea}{\begin{eqnarray}}
\newcommand{\eea}{\end{eqnarray}}
\newcommand{\be}{\begin{equation}}
\newcommand{\ee}{\end{equation}}

\ifams
  \else
 
\fi

\begin{document}

%
\title{
\vspace{1.cm}
A Proof of the Generalized Second Law for Two-Dimensional Black Holes}
\author{Jeongwon Ho\\
\small \vspace{- 3mm}  Department of Physics and Astronomy, University
of Victoria\\
\small \vspace{- 3mm} Victoria, BC, Canada V8W 3P6\\
\small E-mail: {\tt jwho@uvic.ca;@physicist.net}}
\maketitle


\begin{abstract} 
We investigate the generalized second law for two-dimensional black holes in equilibrium (Hartle-Hawking) and nonequilibrium (Unruh) with the heat bath surrounding the black holes. We obtain a simple expression for the change of total entropy in terms of covariant thermodynamic variables, which is valid not only for the Hartle-Hawking state but also for the Unruh state up to leading order, without assuming a quasi-stationary evolution of the black holes. Using this expression, it is shown that the rate of local entropy production is non-negative in the two-dimensional black hole systems. 
\end{abstract}

\newpage


\section{Introduction}

Bekenstein's conjecture of a generalized second law (GSL) for a self-gravitating system \cite{bekenstein73} effectively incapacitated Wheeler's demon, who allows entropy to escape from our universe into black holes and challenges the second law of thermodynamics. According to GSL, a black hole has its own entropy, and then the total entropy $ S_{\rm total}$, which is the sum of the black hole entropy $S_H$ and of the thermodynamic entropy outside the black hole $S_{\rm out}$, never decreases:
\be
\label{gsl}
\Delta S_{\rm total} = \Delta S_H + \Delta S_{\rm out} \geq 0 ~.
\ee

The conjectured GSL was based on the observed relationship between black hole mechanics and ordinary thermodynamics \cite{bardeen73}. In the semiclassical approximation, Hawking has shown that the relationship is not just an analogy. The black hole really emits thermal radiation at a definite temperature like an ordinary thermal object \cite{hawking75}. Moreover, in this analysis, black hole entropy was derived in an exact form as $A/4G$, called the Bekenstein-Hawking entropy, where $A$ is the area of the black hole and $G$ is the gravitational constant. (We adopt units in which $\hbar = c=k=1$.) Since then, our understanding of the microscopic interpretation of the Bekenstein-Hawking entropy has developed considerably, especially in the context of string theory \cite{stringrev}.

However, an explicit general proof of the GSL has not been given until now. When it comes to the question of the fundamentals of the connection between the laws of black hole physics and thermodynamics, the GSL is still an interesting problem; since black hole entropy in the GSL is treated on an equal footing with ordinary thermodynamic entropy, if the GSL holds, a black hole should be nothing but an ordinary thermodynamic object, not just an analogue

Unruh and Wald \cite{unruh82} investigated GSL using gedanken experiments in which one slowly lowers a box containing energy and entropy into a black hole and the reverse of this process. They have shown that taking into account quantum effects, GSL is valid in these processes without introducing Bekenstein's entropy bound. More generally, Frolov and Page \cite{frolov93} proved the validity of GSL for eternal black holes by assuming that the state of matter fields on the past horizon is a thermal state, and that the set of radiation modes on the past horizon and on past null infinity are quantum mechanically uncorrelated. However, these assumptions are not valid for realistic cases such as black holes formed by gravitational collapse. Relaxing these conditions, Mukohyama has shown that the GSL holds for black holes arising from gravitational collapse \cite{muk00}. (See, also \cite{sorkin})

The first law of black hole mechanics is formulated as a relation between infinitesimal variations of the parameters of stationary black holes. In the proofs of GSL in \cite{frolov93}\cite{muk00}, the first law was generalized to a quasi-stationary evolution of the parameters, which means small changes of physical quantities from an initial near-stationary black hole to a final near-stationary black hole. However, the assumption of quasi-stationarity is still only an approximation to the evolution of a black hole formed by gravitational collapse. Apparently, one cannot exclude some amount of net flux of stress-energy induced from vacuum polarization on the (apparent) horizon of the black hole, and then the black hole is not in thermal equilibrium with Hawking radiation surrounding it (Unruh state \cite{unruh}). Thus, in order to investigate the GSL for a black hole arising from gravitational collapse in more realistic situations, the assumption of quasi-stationarity has to be relaxed to a dynamical evolution of thermodynamic variables. Even though the concept of a first law for dynamical black holes is still an open problem, we have a plausible prescription for this developed by Hayward \cite{hayward}. In this paper, we use the term of dynamical first law in Hayward's formulation.

On the other hand, according to relativistic covariance, one can construct a compact form of equation of state with covariant thermodynamic variables such as stress-energy-momentum tensor rather than with energy density and pressure. In particular, when the net energy flux is concerned, i.e., a black hole in a nonequilibrium state, the covariant formulation of black hole thermodynamics has great advantages over ordinary thermodynamic description. (For review of the covariant formulation, see \cite{israel87} and references therein.) 

The purpose of this paper is to prove the GSL for black holes in nonequilibrium states (Unruh) as well as in equilibrium states (Hartle-Hawking \cite{hh}) with Hawking radiation surrounding the black holes. Our strategy is to use the dynamic first law and the covariant formulation of black hole thermodynamics. Then, we obtain a very simple expression for change of total entropy in terms of covariant thermodynamic variables, and it will be seen that the covariant expression for the change of total entropy is valid not only for equilibrium states but also for nonequilibrium states up to leading order.

Our investigation for the GSL will be carried out for a two-dimensional theory of gravity. A proof of GSL for a two-dimensional effective theory of gravity going beyond the assumption of quasi-stationarity was studied in \cite{fiola}. In that paper, the RST version \cite{rst} of CGHS model \cite{cghs} was discussed with the aid of given exact spacetime solutions including the quantum backreaction effects. In this article, an arbitrary two-dimensional static spacetime will be considered for investigation of GSL for the equilibrium states, while for the nonequilibrium states, the S-wave sector in four-dimensional general relativity will be considered.

In Section2, a covariant form of the change of total entropy is obtained for two-dimensional static eternal black holes. Using the stationarity of thermodynamic variables and energy momentum conservation, we show that as expected in a trivial sense, the change of total entropy vanishes and the GSL holds for black holes in equilibrium state. The covariant expression for the change of total entropy of black holes in nonequilibrium states is discussed in Section3. Up to leading order in the Hawking radiation, we show that the change of total entropy is always positive. Section4 will be devoted to a summary of this article.


\section{Hartle-Hawking State}

In the semiclassical approximation, a black hole is surrounded by quantum Hawking radiation which becomes thermal far away from the hole. First, in this section, we assume that a two-dimensional black hole is in thermal equilibrium with the heat bath (Hartle-Hawking state). Our descriptions are given bearing in mind of a stationary eternal black hole with an outer boundary endowed with the Diriclet boundary condition\footnote{The outer boundary with the Dirichlet boundary condition is introduced in a conceptual sense for the black hole to be in equilibrium with the heat bath. Thus, the location of the boundary is out of the problem.}. The backreaction effect of the heat bath is characterized by the expectation value of the renormalized stress-energy tensor of a free massless conformal scalar field, $T_{ab}$. The most general form of static metric satisfying Einstein equations with the source $T_{ab}$ is written by
\begin{equation}
\label{smetric}
ds^2 
     = \frac{dr^2}{f(r)}  - f(r)e^{2\psi (r)} dt^2~,
       ~~~f(r) = 1 - \frac{2 G m(r)}{r}~,
\end{equation}
and the "dilaton" field\footnote{Since the dilaton field does not play an important role in the procedure, we do not require an explicit form of the dilaton field. In addition, our argument given in this section is valid for arbitrary two-dimensional gravitational theory, so the "dilaton" can be considered (for instance) as the scalar field of the effective two-dimensional theory of the S-wave sector of the general relativity and that of the dilaton gravity inspired by the string theory.} $\Phi = \Phi ( r )$. In our system, taking the classical limit in the metric (\ref{smetric}), $\psi (r)$ goes to zero and the mass function $m(r) (> 0) $ becomes a positive constant.

For the procedure, we begin with brief review of local thermodynamic description for the Hartle-Hawking state \cite{muk98}.


\subsection{Local Thermodynamic Description}

In a thermodynamic sense, the Hartle-Hawking state is described by local equations of state, so called Duhem-Gibbs relations,
\begin{equation}
\label{dgrelations}
s 
   = \beta (\rho + P)~,
      ~~~ ds =\beta d\rho ~,
\end{equation}
where $\beta (r)^{-1} = T (r)$ is a local temperature parameterizing the equilibrium state and $s(r)$, $\rho (r) = - T^t_t$, $P(r) = T^r_r $ are local thermodynamic variables identified as entropy density, quantum energy density, and pressure, respectively. Using the Duhem-Gibbs relations together with the conservation law $ \nabla_a T^a_b = 0 $, one finds that distribution of the stress-energy also satisfies the Tolman condition
\begin{equation}
\label{tcond}
                     \sqrt{-g_{tt}}T(r) = {\rm constant}~.
\end{equation}

On the other hand, the local thermodynamic variables are to be completely determined (up to a boundary condition) by the conservation law and the trace anomaly $ T^a_a = R/ 24\pi $;
\begin{eqnarray}
\label{pressure}
P(r) &=& \frac{1}{24\pi} \frac{\kappa_H^2 - \kappa^2}{f e^{2\psi}}~,
\\
\label{density}
\rho (r) &=& \frac{1}{24 \pi} \left( 2 e^{- \psi} \frac{d
\kappa}{d r} + \frac{ \kappa^2_H - \kappa^2 }{ f e^{2 \psi}} \right)~,
\\
\label{endensity}
s(r) &=&  \frac{1}{6 \kappa_H} \left( \sqrt{f} \frac{d
\kappa}{d r} + \frac{ \kappa^2_H - \kappa^2 }{ \sqrt{f} e^{\psi}} \right)~,
\end{eqnarray}
where $\kappa$ is the surface gravity calculated with respect to a timelike Killing vector $\xi^a = \partial x^a / \partial t$ given as
\begin{equation}
\label{sfgravityhh}
\kappa = e^{\psi } \left( \frac{1}{2} \partial_r f + f \partial_r \psi \right)
\end{equation}
and $\kappa_H =  \kappa (r_H)$ is the surface gravity at the Killing horizon $ r =r_H$ determined by $f(r_H) =0$. The integration constant in (\ref{pressure}) and (\ref{density}) was chosen so that the energy density and pressure are regular at the horizon.

From the equation (\ref{endensity}), entropy outside the black hole is given by
\be
\label{outentropy1}
       S_{\rm out} =  \int^{r_0}_{r_H} dr \sqrt{g_{rr}} s(r)~,
\ee
where $r_0$ is an outer boundary. In fact, the outside entropy given by (\ref{outentropy1}) is equivalent to the fine-grained entropy evaluated in \cite{fiola}; the infrared divergent part of $S_{\rm out}$ becomes \cite{israelpv}
\begin{equation}
\label{irentropy}
S^R_{\rm out} \approx \frac{1}{6} \kappa_H (r_0 - r_H )~,
\end{equation}
where $r_0 \rightarrow \infty $. Then $S^R_{\rm out}$ precisely matches the infrared divergent part of the fine-grained entropy given in (95) in \cite{fiola}. It is quite interesting to notice that while the fine-grained entropy suffers from the typical ultraviolet divergence appearing near horizon and requires a cutoff near the horizon, the outside entropy $S_{\rm out}$ is free of ultraviolet divergences; according to the definition of the Hartle-Hawking state, the thermodynamic variables in (\ref{pressure})-(\ref{endensity}) are regular at the horizon.


\subsection{Covariant Formulation and GSL}

It is well known that the Gibbs laws (\ref{dgrelations}) can be translated into covariant equations given by
\be
\label{dgco1}
s^a = P \beta^a -\beta^b T^a_b~~,
d s^a =  - \beta^b d T^a_b~,
\ee
where $\beta^a = \beta_H \xi^a = \beta u^a$, $u^a u_a =-1$, and $\beta_H = T_H^{-1}$ is the inverse Hawking temperature. The entropy density vector $s^a$ is defined by $s^a = s(r)u^a$ and $T^a_b$ in terms of the energy density and pressure is written by
\be
\label{setensorst}
T_a^b = \rho u_a u^b + P \Delta_a^b ~,
\ee
where $\Delta_a^b = g_a^b + u_a u^b$.

Now, consider an evolution of two chosen spacelike hypersurfaces ($\Sigma_1$, $\Sigma_2$)\footnote{Through out the paper, it is understood that the spacelike hypersurfaces are parameterized with a time function that well behaves at the horizon such as Kruskal-like time coordinate.}. The total entropy change in the evolution is composed of the sum of the change of black hole entropy $\Delta S_H$ plus the change of entropy outside the black hole $\Delta S_{\rm out}$,
\be
\label{totentropy}
\Delta S_{\rm total} = \Delta S_H + \Delta S_{\rm out}~.
\ee
In terms of covariant variables, the entropy change of outside black hole is
\be
\label{enchhy}
\Delta S_{\rm out} = S_{\Sigma_2} - S_{\Sigma_1}
                   = \int^{\Sigma_2}_{\Sigma_1} s^a d \hat{\Sigma}_a ~,
\ee
where $\int^{\Sigma_2}_{\Sigma_1}$ means integration over $\Sigma_2$ minus integration over $\Sigma_1$. The second equality in (\ref{enchhy}) can be easily checked comparing with (\ref{outentropy1}) as following;
\bea
\label{outentropy}
S_{\rm out} &=& \int_{\Sigma} s^a d \hat{\Sigma}_a
\nonumber \\ 
            &=& \beta_H \int^{r_0}_{r_H} d r e^{\psi (r)} \left(P (r) +
                    \rho (r) \right)
\nonumber \\
            &=& \int^{r_0}_{r_H} dr \sqrt{g_{rr}} s(r)~.
\eea

On the other hand, the change of the black hole entropy should arise from heat flow going through the horizon. To be exact, heat passing through the horizon causes an entropy change of the black hole and work done along the horizon (in the case of non-vanishing net flux). Obviously, work term is not generated on the Killing horizon with vanishing net flux. Thus, the change of black hole entropy should be equivalent to the amount of heat flow passing through the horizon
\be
\label{enchbh}
\Delta S_H = S_{{\cal H}_2} - S_{{\cal H}_1}  
           = \int^{{\cal H}_2}_{{\cal H}_1} s^a d{\hat {\cal H}}_a ~,
\ee
where ${\cal H}_i$ are intersecting points of the horizon ${\cal H}$ and $\Sigma_i$.

Substituting (\ref{enchhy}), (\ref{enchbh}) into (\ref{totentropy}), and using Gauss-Stokes theorem, we obtain a simple form of change of total entropy given by
\bea
\label{chtoten}
\Delta S_{\rm total}
       &=& \int^{{\cal H}_2}_{{\cal H}_1} s^a d{\hat {\cal H}}_a
      +  \int^{\Sigma_2}_{\Sigma_1} s^a d \hat{\Sigma}_a
\nonumber \\
       &=& \int_M d^2x \sqrt{-g} \nabla_a s^a  ~,
\eea
where $M$ denotes a spacetime patch\footnote{Strictly speaking, the spacetime patch $M$ is also bounded by the outer boundary. However, since the Dirichlet boundary condition is imposed on the boundary, the surface integral on the boundary $\int s^a d \hat{l}_a $ does not give any effect to the change of total entropy given by (\ref{chtoten})} bounded by the spacelike hypersurfaces $\Sigma_1$, $\Sigma_2$, and the horizon ${\cal H}$. Thus, GSL for the Hartle-Hawking state can be easily estimated by determining the sign of the term $\nabla_a s^a$  ($ \int_M d^2x \sqrt{-g} \nabla_a s^a $) in a strong (weak) sense; hereafter, we refer $\nabla_a s^a$ to rate of local entropy production.

Using the conservation law of the stress-energy tensor and the Killing equation $\beta_{(a;b)} =0$, we find that the density of total entropy change vanishes $\nabla_a s^a = 0$. Thus, {\it GSL holds for any two-dimensional eternal black holes in equilibrium with surrounding Hawking radiation in a `strong' sense}. In fact, GSL for the Hartle-Hawking state is trivially satisfied, because according to the inherence of the Hartle-Hawking state, on the Killing horizon (and the outer boundary), in-flux exactly compensates for out-flux.

In \cite{somi} Shimomura {\it et al} investigated the GSL for two-dimensional black holes, and argued that if one takes a finite region outside the black hole, GSL does not hold even for the Hartle-Hawking state. As argued in \cite{somi}, the violation of GSL is caused by fixing the outer boundary of the finite accessible region so that the size of the accessible region decreases. However, fixing the outer boundary, one should miss in-flux from infinity to the finite accessible region passing through the outer boundary, and the violation of GSL in \cite{somi} is an artificial effect. In this paper, we have required that the outer boundary evolve satisfying the Dirichlet boundary condition and maintaining the balance between in-flux and out-flux on the boundary rather than fixing the outer boundary.


\section{Unruh State}


\subsection{S-wave Sector}

Now, consider a black hole out of equilibrium with the thermal heat bath (Unruh state). Our two-dimensional black holes considered here are S-wave sector in four-dimensional general relativity. The action is given by spherically symmetric reduction;
\begin{eqnarray}
\label{gaction}
I_G &=& \frac{1}{16 \pi G}\int d^4x \sqrt{-g^{(4)}} R^{(4)} + {\rm
surface~terms} \nonumber \\
    &=& \frac{1}{4G} \int d^2 x \sqrt{-g} ( r^2 R + 2 ( \nabla r )^2 +
2) + {\rm surface~terms}~,
\end{eqnarray}
where four-dimensional spherically symmetric spacetime metric
\be
\label{fmetric}
ds^{2}_{(4)} = g_{ab} dx^a dx^b + r^2(x)d \Omega^2~,
\ee
was used, and the covariant derivative is defined with respect to the two-dimensional metric $g_{ab}$. Here, we refer two-dimensional geometric quantities without dimensional notations, such as superscript $^{(4)}$.

On the other hand, taking into account the Hawking radiation and its backreaction on the spacetime in the semiclassical approximation, one-loop quantum effective action is to be added to the classical gravitational action (\ref{gaction}),
\be
\label{action}
I = I_G + \Gamma~.
\ee
In a self-consistent manner, the two-dimensional one-loop effective action has to be obtained by the same spherically symmetric reduction of four-dimensional matter fields as has been done for the gravitational part $I_G$. In our analysis, however, we assume that the effective two-dimensional matter is conformal and the expectation value of the stress-energy tensor defined by $T_{ab} = (-2/\sqrt{-g})(\delta \Gamma / \delta g^{ab})$ is given as an Unruh-like one
\be
\label{setensor}
T_a^b = \rho u_a u^b + P \Delta_a^b - F_0 l_a l^b,
\ee
where $l_a$ is an affinely parameterized ingoing light-like vector, and $F_0$ denotes the Hawking flux.

Variations of the action $I$ with respect to the two-dimensional metric $g_{ab}$ and `dilaton' $r^2$ give field equations as follows
\bea
\label{einseq}
& G_{ab} \equiv
-2 r \nabla_a \nabla_b r + g_{ab}(2 r \Box r + (\nabla r)^2 -1)
= 2 G T_{ab}~, &
\nonumber \\
& r R - 2 \Box r =0~. &
\eea
A general solution to the field equations (\ref{einseq}) can be written by the type considered by Bardeen \cite{bardeen},
\bea
\label{tmetric}
 ds^2 &=& - f(v,r)e^{2 \psi (v,r)}dv^2 + 2 e^{\psi (v,r)}dvdr~,
\nonumber \\
f(v,r) &=& 1 - \frac{2Gm(v,r)}{r}~,
\eea
which becomes the Vaidya metric for $\psi = 0$ and $m=m(v)$, and the Schwarzshild metric in Eddington-Finkelstein coordinates for $\psi = 0$ and $m = {\rm constant} > 0$. The apparent horizon, $r_H $, will be given by the relation $ f(r=r_H, v) =0$. Substituting (\ref{tmetric}) into (\ref{einseq}), then the Einstein's equations become
\be
\label{simple}
\frac{\partial m}{\partial r} = \rho,~~
\frac{\partial m}{\partial v} = - e^{- \psi} F_0,~~
\frac{\partial \psi}{\partial r} = \frac{G}{r}f^{-1}(P + \rho)~.
\ee

In order to formulate the black hole thermodynamics on a non-static spacetime such as (\ref{tmetric}), one needs appropriate definitions of energy and surface gravity in a dynamical sense. The first law of black hole dynamics proposed in \cite{hayward} was given in terms of Misner-Sharp energy \cite{misner64}, $m(v,r)$ in our coordinates system,
\be
\label{msenergy}
m(r,v) = \frac{r}{2G} \left( 1 - (\nabla r)^2 \right)~,
\ee
and dynamical surface gravity $\kappa^{dyn} $
\be
\label{sufgravity}
\kappa^{dyn} = \frac{1}{2} \Box r = \frac{1}{2} \left( \partial_r f + f \partial_r \psi \right)~.
\ee
The dynamical surface gravity (\ref{sufgravity}) is evaluated with respect to the 1-form $k = * {\bf d}r$ \cite{kodama80}, where ${\bf d}$ is the exterior derivative and $*$ is the Hodge operator, which generates a preferred flow of time and is a dynamic analogue of the stationary Killing vector \cite{hayward}. Then, the dynamical first law is obtained from Einstein equations and holds for any apparent horizon\footnote{In fact, apparent horizon might be conceptually different from trapping horizon, which is the term used in Hayward's articles for local definition of a dynamical black hole. Since, in our analysis, there is not any reason to distinguish between them, we just use the term apparent horizon.}. In the case of two-dimensional black holes, it is written as
\be
\label{firstlaw}
\beta^{dyn}_H \nabla^\prime m_H = \nabla^\prime S_H +
                   \beta^{dyn}_H\omega_H \nabla^\prime r ~,
\ee
where $\nabla^\prime $ denotes the derivative along the apparent horizon and $\omega \equiv -T^a_a/2 $. $(\beta^{dyn}_H)^{-1} =T^{dyn}_H =\kappa^{dyn}/2 \pi $ is the inverse dynamical Hawking temperature. Note that the second term on the right hand side of (\ref{firstlaw}) applies to work done along the horizon.

On the other hand, the dynamical surface gravity $\kappa^{dyn}$ defined in (\ref{sufgravity}) is obtained in a remarkably simple form by using the equations (\ref{einseq}) and the trace anomaly; The trace anomaly together with the dilaton equation (\ref{einseq}) is written as
\be
\label{trace1}
T^a_a = \frac{R}{24 \pi} = \frac{\kappa^{dyn}}{6 \pi r}~.
\ee
Using the Einstein equations (\ref{einseq}), the trace of the stress-energy tensor is also given by 
\be
\label{trace2}
T^a_a = \frac{1}{G} \left(2r \kappa + f -1 \right)~.
\ee
Then, using equations (\ref{trace1}) and (\ref{trace2}), we obtain the dynamical surface gravity
\be
\label{sufgravity1}
\kappa^{dyn}
= \frac{Gm}{r^2} \left(1 - \frac{G}{12 \pi r^2} \right)^{-1}~,
\ee
and dynamical temperature on the horizon
\be
\label{sufgravity2}
T^{dyn}_H = \frac{\kappa^{dyn}_H}{2 \pi} =
    \frac{1}{ 8 \pi G m_H} \left( 1 - \frac{
             1}{48 \pi G m_H^2} \right)^{-1} ~,
\ee
where $m_H = m(r_H, v)$. It is quite interesting that the one-loop correction to dynamical temperature appeared in (\ref{sufgravity2}) has exactly the same form as the one-loop correction to the Hawking temperature of a static black hole evaluated in \cite{fursaev}. However, it has to be noticed that since in our analysis $m_H$ is not a constant, but contains quantum corrections, the factor\ $1/ 8\pi G m_H $ cannot be interpreted as the `classical' Hawking temperature. Instead, quantum corrections to the dynamical temperature are contained in the factor $1/ 8\pi G m_H $ as well as in the factor in parentheses of (\ref{sufgravity2}).


\subsection{GSL for Unruh State}

Now, we are ready to examine the GSL for the Unruh state. First, it is obvious that the expression for the outside entropy change in the Hartle-Hawking state given in (\ref{enchhy}) is still valid for the Unruh state. However, for the Unruh state, the expression for the black hole entropy change given in (\ref{enchbh}) should be corrected by a work term, which is generated along the apparent horizon by heat flow passing through the horizon.  Let us evaluate the integral in (\ref{enchbh});
\bea
\label{enchbh1}
\int^{{\cal H}_2}_{{\cal H}_1} s^a d{\hat {\cal H}}_a 
 &=& \int_{{\cal H}_1}^{{\cal H}_2}
            e^{\psi_H}   \left( s^v dr - s^r dv \right)_{\cal H}
\nonumber \\
 &=& \int_{{\cal H}_1}^{{\cal H}_2} \beta^{dyn}_H dm_H
\nonumber \\
 &=& 4\pi G \left(m_{H_2}^2 - m_{H_1}^2 \right)
   - \frac{1}{6} \log \left( \frac{m_{H_2}}{m_{H_1}} \right).
\eea
As well known, a typical one-loop correction to black hole entropy is given by a logarithmic term \cite{fursaev} and the quantity given in (\ref{enchbh1}) looks like just the change of (one-loop corrected) black hole entropy. However, this is not true. It has to be noticed that from the viewpoint of the black hole, the heat flow passing through the horizon causes the work done along the horizon as well as the change of black hole entropy, and the logarithmic term represents the work along the horizon from ${\cal H}_1$ to ${\cal H}_2$. It could be checked from the dynamical first law (\ref{firstlaw}); performing integrations along the horizon on the both side of the first law, one finds that on the right hand side, the first term gives the Bekenstein-Hawking entropy (area law) and the second term, which is understood as the work term, becomes the logarithmic term appearing in (\ref{enchbh1}).

Thus, the volume integration of the divergence of the entropy vector (\ref{chtoten}) represents the sum of the change of total entropy and of the work done along the horizon. However, as we have shown in the above paragraph, fortunately, the work term contributes to the integration in only subleading order. So, up to leading order, the change of total entropy for the Unruh state is still given by the volume integral (\ref{chtoten}), just like the case of Hartle-Hawking state.

Using (\ref{simple}) and (\ref{sufgravity1}), the rate of local entropy production becomes
\bea
\label{totenchun}
\nabla_a s^a 
 &=&
   \frac{4 \pi r}{G m} F_0 e^{-2\psi} \left(
1 + \frac{r}{2} \left(
\partial_r \psi + \partial_r \ln m \right)\left(
1 - \frac{G}{12 \pi r^2} \right) \right)
\nonumber \\
& \approx & \frac{4 \pi r}{G m} F_0 e^{-2\psi}~,
\eea
where approximation has been made for taking only leading term in the evaluation. Finally, we find that {\it the rate of local entropy production for the Unruh state is always positive in leading order.} Note that taking a static limit of the two-dimensional spacetime is taken, $F_0 = 0$, the rate of local entropy production vanishes, and we return to the case of the Hartle-Hawking state. In addition, the rate of local entropy change calculated in the equation (\ref{totenchun}) can be also obtained by considering only near the horizon.

\section{Concluding Remarks}

Our main result in this article was to obtain the covariant form of the change of total entropy given by (\ref{chtoten}) and verify the GSL for two-dimensional black holes in equilibrium and nonequilibrium with the heat bath. Especially, we were interested in the dynamical evolution of the black holes and its contribution to the change of total entropy.

For two-dimensional (eternal) black holes in thermal equilibrium with the heat bath, we have shown that the rate of local entropy production $\nabla_a s^a$ vanishes and the GSL holds for the system. That the rate of local entropy production vanishes in the equilibrium situation is due to the balance between in-flux and out-flux on the Killing horizon and the outer boundary with the Dirichlet boundary condition. On the other hand, it has been argued that the covariant form of the change of total entropy is still valid for the nonequilibrium situation up to leading order. We have shown that the rate of local entropy production for the nonequilibrium system is positive at the leading order.

Our important tools for the investigation was the covariant formulation and the dynamical first law (\ref{firstlaw}) of black hole thermodynamics. In fact, there may be a criticism of the dynamical black hole thermodynamics introduced by Hayward. However, the dynamical first law given in equation (\ref{firstlaw}) is nothing but the Einstein equation on the horizon and in \cite{hma} the dynamic black hole entropy in the first law has been inspected by using the original Clausius definition of thermodynamic entropy and the Wald's Noether charge method \cite{wald}. Even though the definition of the dynamical temperature in the dynamical black hole thermodynamics can be flawed by considering general physical systems, since in this article the source term of the r.h.s. of Einstein equations is contributed by only the conformal scalar field, which effectively characterizes the quantum backreaction, not by any other ordinary matter fields, the dynamical temperature is well defined. Indeed, we obtained the dynamical temperature on the horizon with the one-loop correction (\ref{sufgravity2}), which has exactly the same form as the one-loop correction to the Hawking temperature of a static black hole, and it can be seen that taking the classical limit (excluding the quantum backreaction), it recovers the Hawking temperature of a static spacetime.

\section*{Acknowledgment}
I thank W. Israel and S. Mukohyama for very helpful discussions and suggestions to this work.


\end{document}